\documentclass[aps,twocolumn,superscriptaddress]{revtex4}
\usepackage{amsmath}
\usepackage{graphicx}

\begin{document}
\bibliographystyle{apsrev}

\title{Spin-Orbit Coupling, Antilocalization, and Parallel Magnetic
Fields in Quantum Dots}

\author{D. M. Zumb\"uhl}
\affiliation{Department of Physics, Harvard University, Cambridge,
Massachusetts 02138}

\author{J. B. Miller}
\affiliation{Department of Physics, Harvard University, Cambridge,
Massachusetts 02138}
\affiliation{Division of Engineering and Applied Sciences, Harvard
University, Cambridge,
Massachusetts 02138}

\author{C. M. Marcus}
\affiliation{Department of Physics, Harvard University, Cambridge,
Massachusetts 02138}
\author{K. Campman}
\affiliation{Department of Electrical and Computer Engineering,
University of California, Santa
Barbara, California 93106}
\author{ A. C. Gossard}
\affiliation{Department of Electrical and Computer Engineering,
University of California, Santa
Barbara, California 93106}


\begin{abstract} We investigate antilocalization due to spin-orbit coupling in ballistic GaAs quantum dots. Antilocalization
that is prominent in large dots is suppressed in small dots, as anticipated theoretically. Parallel magnetic fields
suppress both antilocalization and also, at larger fields, weak localization, consistent with random matrix theory
results once orbital coupling of the parallel field is included. \emph{In situ} control of spin-orbit coupling in dots
is demonstrated as a gate-controlled crossover from weak localization to antilocalization.
\end{abstract}

\maketitle

The combination of quantum coherence and electron spin rotation in mesoscopic systems produces a number of interesting
and novel transport properties. Numerous proposals for potentially revolutionary electronic devices that use spin-orbit
(SO) coupling have appeared in recent years, including gate-controlled spin rotators \cite{DattaDas} as well as sources
and detectors of spin-polarized currents \cite{SOPol}. It has been predicted that the effects of some types of  SO
coupling will be strongly suppressed in small 0D systems, i.e., quantum dots \cite{Khaetskii,Halperin,RMT}. This
suppression as well as overall control of SO coupling will be important if quantum dots are used to store electron spin
states as part of a future information processing scheme.

In this Letter, we investigate SO effects in ballistic-chaotic GaAs/AlGaAs quantum dots. We identify the signature of
SO coupling in ballistic quantum dots to be {\em antilocalization} (AL), leading to characteristic magnetoconductance
curves, analogous to known cases of disordered 1D and 2D systems
\cite{HLK,BergmannReview,DresselhausExp,Millo,Miller,Knap}. AL is found to be prominent in large dots and suppressed in
smaller dots, as anticipated theoretically \cite{Khaetskii,Halperin,RMT}. Results are generally in excellent agreement
with a new random matrix theory (RMT) that includes SO and Zeeman coupling \cite{RMT}. Moderate magnetic fields applied
in the plane of the 2D electron gas (2DEG) in which the dots are formed cause a crossover from AL to weak localization
(WL). This can be understood as a result of Zeeman splitting, consistent with RMT \cite{RMT}. At larger parallel fields
WL is also suppressed, which is not expected within RMT. The suppression of WL is explained quantitatively by orbital
coupling of the parallel field, which breaks time-reversal symmetry \cite{Falko}. Finally, we demonstrate \emph{in
situ} electrostatic control of the SO coupling strength by tuning from AL to WL in a dot with a center gate.

It is well known that in mesoscopic samples coherent backscattering of time-reversed electron trajectories leads to a
conductance {\em minimum} (WL) at $B=0$ in the spin-invariant case, and a conductance {\em maximum} (AL) in the case of
strong SO coupling \cite{HLK}. In semiconductor heterostructures, SO coupling results mainly from electric fields
\cite{DyakanovPerel} (appearing as magnetic fields in the electron frame) leading to momentum dependent spin
precessions due to crystal inversion asymmetry (Dresselhaus term \cite{Dresselhaus}) and heterointerface asymmetry
(Rashba term \cite{Rashba}).

SO coupling effects have been previously measured using AL in GaAs 2DEGs \cite{DresselhausExp, Millo, Miller} and other
2D heterostructures \cite{Knap}. Other means of measuring SO coupling in heterostructures, such as from Shubnikov-de
Haas oscillations \cite{SdH} and Raman scattering spectroscopy \cite{Jusserand} are also quite developed. SO effects
have also been reported in mesoscopic systems (comparable in size to the phase coherence length) such as Aharonov-Bohm
rings, wires, and carbon nanotubes \cite{rwt}. Recently, parallel field effects of SO coupling in quantum dots were
measured \cite{Hackens, Folk}. In particular, an observed reduction of conductance fluctuations in a parallel field
\cite{Folk} was explained by including SO effects \cite{Halperin, RMT}, leading to an important extension of random
matrix theory (RMT) to include new symmetry classes associated with SO and Zeeman coupling \cite{RMT}.

\begin{figure}[b]
             \label{fig1}
      \vspace{-0.2in}
\includegraphics{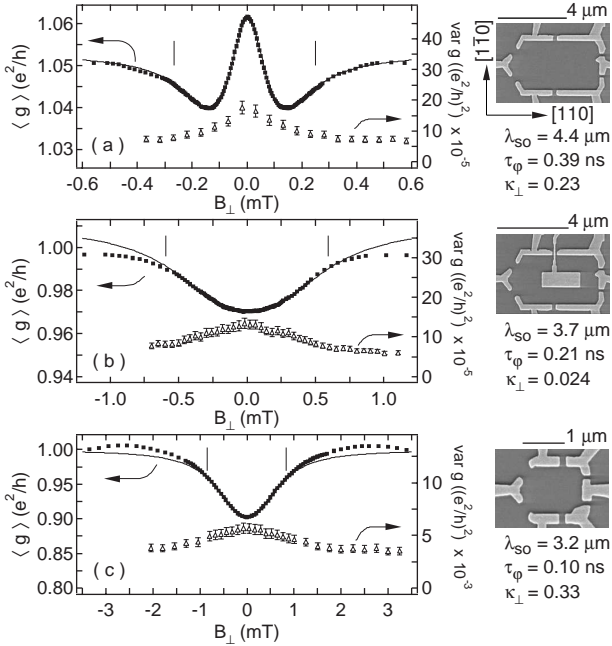}
      \vspace{-0.1in}
       \caption{\footnotesize {Average conductance $\langle g \rangle$
(squares) and variance of
       conductance $\mathrm{var}(g)$ (triangles) calculated from $\sim 200$
statistically independent samples
       (see text) as a function of perpendicular magnetic field
$B_\bot$ for (a) $8.0\,\mathrm{\mu m^2}$ dot (b)
       $5.8\,\mathrm{\mu m^2}$ center-gated dot and (c)
$1.2\,\mathrm{\mu m^2}$ dot at $T = 0.3\,\mathrm{K}$, along with fits to RMT (solid curves).
       In (b), the center gate is fully depleted. Vertical lines
       indicate the fitting range, error bars of $\langle g \rangle$
are about the size of the squares.}}
\end{figure}

This RMT addresses quantum dots coupled to two reservoirs via $N$ total conducting channels, with $N\gg 1$. It assumes
$(\gamma, \epsilon_Z)\ll E_{T}$, where $\gamma =N\Delta/(2\pi)$ is the level broadening due to escape, $\Delta$ is the
mean level spacing, $\epsilon_{Z}=g\mu_{B} B$ is the Zeeman energy and $E_{T}$ is the Thouless energy (Table I).
Decoherence is included as a fictitious voltage probe \cite{BBB, RMT} with dimensionless dephasing rate $N_\varphi =
h/(\Delta\tau_\varphi)$, where $\tau_\varphi$ is the phase coherence time. SO lengths $\lambda_{1,2}$ along respective
principal axes $\lbrack 110 \rbrack$ and $\lbrack 1 \bar 10 \rbrack$ are assumed (within the RMT) to be large compared
to the dot dimensions $L_{1,2}$ along these axes. We define the mean SO length $\lambda_{so} = \sqrt{|\lambda_1
\lambda_2|}$ and SO anisotropy $\nu_{so}=\sqrt{|\lambda_1 / \lambda_2|}$. SO coupling introduces two energy scales:
$\epsilon_{\bot}^{so} = \kappa_\bot E_T (L_1 L_2/\lambda_{so}^2)^2$, which represents a spin-dependent
Aharonov-Bohm-like effect, and $\epsilon_\parallel^{so} \sim ((L_1/\lambda_1)^2+(L_2/\lambda_2)^2) \epsilon_\bot^{so}$,
providing spin flips. AL appears in the regime of strong SO coupling, $(\epsilon_{\bot}^{so}, \epsilon_\parallel^{so})
\gg \tilde \gamma$, where $\tilde \gamma$ is the total level broadening $\tilde \gamma = ( \gamma +
\hbar/\tau_\varphi)$.  Note that large dots reach the strong SO regime more readily (i.e., for weaker SO coupling) than
small dots. Parameters $\lambda_{so}$, $\tau_\varphi$, and $\kappa_\bot$ (a dimensionless parameter characterizing
trajectory areas within the dot) are extracted from fits to dot conductance as a function of perpendicular field,
$B_\perp$. The asymmetry parameter, $\nu_{so}$, is estimated from the dependence of magnetoconductance on parallel
field, $B_\parallel$.

The quantum dots are formed by lateral Cr-Au depletion gates defined by electron-beam lithography on the surface of a
GaAs/AlGaAs heterostructure grown in the [001] direction. The 2DEG interface is $349\, \mathrm{\AA}$ below the wafer
surface, comprising a $50\, \mathrm{\AA}$ GaAs cap layer and a $299\, \mathrm{\AA}$ AlGaAs layer with two Si
$\mathrm{\delta}$-doping layers $143\, \mathrm{\AA}$ and $161\, \mathrm{\AA}$ from the 2DEG. An electron density of
$n\sim 5.8\times 10^{15}\, \mathrm{m^{-2}}$ \cite{Subb} and bulk mobility $\mu \sim24\, \mathrm{m^2/Vs}$ (cooled in the
dark) gives a transport mean free path $\ell_{e}\sim3\, \mathrm{\mu m}$. This 2DEG is known to show AL in 2D
\cite{Miller}. Measurements were made in a $ ^3$He cryostat at $0.3\, \mathrm{K}$ using current bias of $1 \,
\mathrm{nA}$ at $338\,\mathrm{Hz}$. Shape-distorting gates were used to obtain ensembles of statistically independent
conductance measurements \cite{Chan95} while the point contacts were actively held at one fully transmitting mode each
($N=2$).

\begin{table}[b]
      \vspace{-0.2in}
            \label{table1}
               \begin{tabular}{c|c|c|c|c|c|c|cc}
               \rule[-2mm]{0mm}{3mm} A & $\Delta$
               & $\tau_d$ & $E_{T}/\Delta$ & $\epsilon_\bot^{so}/\Delta$
               & $\epsilon_\parallel^{so}/\Delta$ & $a_1$, $a_2$ & $b_2$ \\
               $\, \mathrm{\mu m^2}$& $\mathrm{\mu eV}$ & ns
&&&&$\mathrm{(ns)^{-1}T^{-2}}$\hspace{0.3cm}& $\mathrm{(ns)^{-1}T^{-6}}$ \\
\hline
               1.2 &  6.0 & 0.35 &  33 & 0.15 & 0.04 & 6.6, 6.6 & 0.24 \\
               5.8 &  1.2 & 1.7 & 73 & 0.32 & 0.33 & 3.2, 0   & 140  \\
               8   &  0.9 & 2.3 & 86 & 3.6 & 3.1 & 1.4, 0.9 & 3.7  \\
               \end{tabular}
              \caption{\footnotesize{Dot area $A=L_1 L_2$ ($ 130\,
\mathrm{nm}$ edge depletion);
              spin-degenerate mean level spacing $\Delta =
2\pi\hbar^2/m^*A$ ($m^*=0.067m_e$);
              dwell time $\tau_d=h/(N \Delta)$;
              Thouless energy $E_{T} = \hbar v_F / \sqrt{A}$;
              $\epsilon_{\bot}^{so}/\Delta$ and $\epsilon_\parallel^{so}/\Delta$
               for the fits in Fig.\ 1;
              $B^2$ coefficients $a_1$ and $a_2$ from one and two parameter fits;
              $B^6$ coefficient $b_2$ from two parameter fit, see text.
              }}
\end{table}

Figure 1 shows average conductance $\langle g \rangle$, and variance of conductance fluctuations, $\mathrm{var}(g)$, as
a function of $B_\perp$ for the three measured dots: a large dot ($A \sim 8\, \mathrm{\mu m^2}$), a variable size dot
with an internal gate ($A \sim 5.8\,\mathrm{\mu m^2}$ or $8\,\mathrm{\mu m^2}$, depending on center gate voltage), and
a smaller dot ($1.2\, \mathrm{\mu m^2}$). Each data point represents $\sim200$ independent device shapes. The large dot
shows AL while the small and gated dots show WL.  Estimates for $\lambda_{so}$, $\tau_{\varphi}$ and $\kappa_\bot$,
from RMT fits are listed for each device below the micrographs in Fig.~1 (see Table I for corresponding $\epsilon_\bot$
and $\epsilon_\parallel$). When AL is present (i.e., for the large dot), estimates for $\lambda_{so}$ have small
uncertainties ($\pm5\%$) and give upper and lower bounds; when AL is absent (i.e., for the small and gated dots) only a
lower bound for $\lambda_{so}$ ($-5\%$) can be extracted from fits.  The value $\lambda_{so} \sim 4.4\,\mathrm{\mu m}$
is consistent with all dots and in good agreement with AL measurements made on an unpatterned 2DEG sample from the same
wafer \cite{Miller}.

\begin{figure}[t]
             \label{figp2}
             \includegraphics{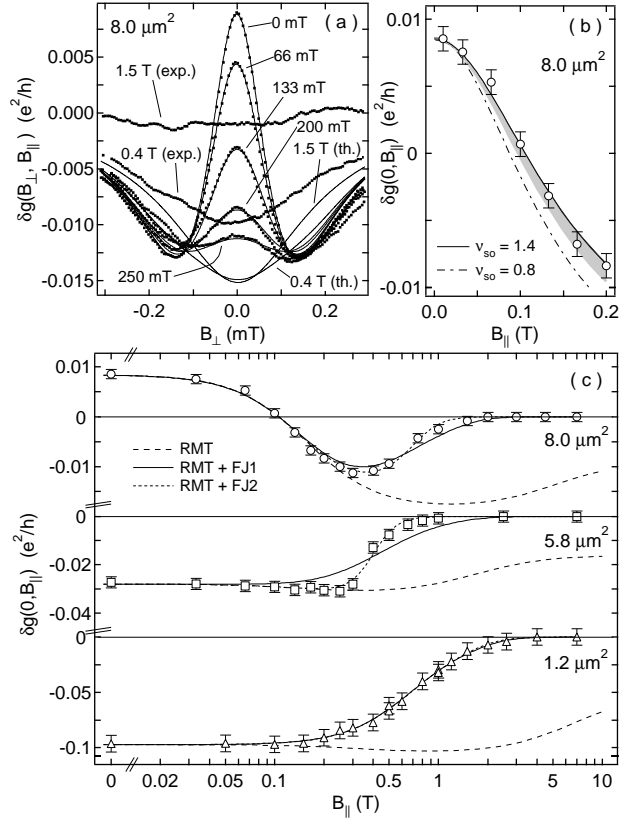}
     \vspace{-0.35in}
     \caption{\footnotesize{(a) Difference of average conductance from
its value at large $B_\bot$, $\delta g(B_\bot,B_\parallel)$, as a function of $B_{\bot}$ for several $B_\parallel$ for
the $8.0\,\mathrm{\mu m^2}$ dot at $T = 0.3\,\mathrm{K}$ (squares) with RMT fits (curves).
      (b) Sensitivity of $\delta g (0,B_\parallel)$ to $\nu_{so}$ for
the $8.0\,\mathrm{\mu m^2}$ dot, $1 \leq \nu_{so} \leq
      2$ (shaded), $\nu_{so}=1.4$ (solid line) and $\nu_{so}=0.8$ (dashed line)
      (c) $\delta g(0,B_\parallel)$ (markers) with RMT predictions
(dashed curves) and one parameter (solid curves) or two parameter fits
      (dotted curves) using RMT including a suppression
factor due to orbital coupling of $B_\parallel$, see text.
     \vspace{-0.225in}
        }}
\end{figure}

Comparing Figs.~1(a) and 1(c), and recalling that all dots are fabricated on the same wafer, one sees that AL is
suppressed in smaller dots, even though $\lambda_{so}$ is sufficient to produce AL in the larger dot. We note that
these dots do not strongly satisfy the inequalities $L/\lambda_{so} \ll1, N\gg 1$, having $N=2$ and $L/\lambda_{so} =
0.64\ (0.34)$ for the large (small) dot. Nevertheless, Fig.~1 shows the very good agreement between experiment and the
new RMT.

We next consider the influence of a parallel magnetic field on average magnetoconductance. In order to apply
tesla-scale $B_\parallel$ while maintaining subgauss control of $B_\perp$, we mount the sample with the 2DEG aligned to
the axis of the primary solenoid (accurate to $\sim 1^\circ$) and use an independent split-coil magnet attached to the
cryostat to provide $B_\perp$ as well as to compensate for sample misalignment \cite{Folk}. Figure 2 shows plots of the
deviation of the shape-averaged conductance from its value at $B_\perp \gg \phi_0/A$ (i.e., with time-reversal symmetry
fully broken by $B_\perp$), $\delta g(B_\bot,B_\parallel)=\langle g(B_\bot,B_\parallel) \rangle - \langle
g(B_\perp\gg\phi_0 / A ,B_\parallel) \rangle$. Figure 2(a) shows $\delta g(B_\bot,B_\parallel)$ as a function of
$B_\bot$ at several values of $B_\parallel$, along with fits of RMT \cite{RMT} in which parameters $\lambda_{so}$,
$\tau_\varphi$ and $\kappa_\bot$ have been set by a single fit to the $B_\parallel = 0$ data. The low-field dependence
of $\delta g(0,B_\parallel)$ on $B_\parallel$ (Fig.~2(b)) then allows the remaining parameter, $\nu_{so}$, to be
estimated as described below.

\begin{figure}[t]
             \label{figp3.epsf}
             \includegraphics{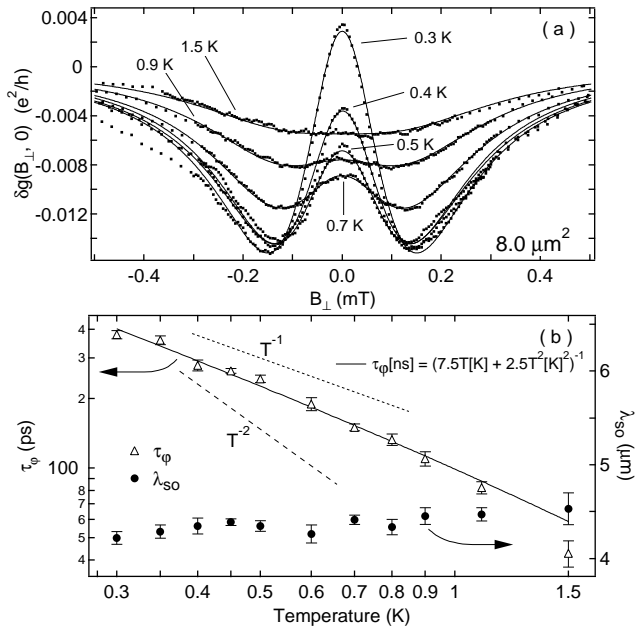}
      \vspace{-0.3in}
      \caption{\footnotesize{(a) Difference of average conductance from
its value at large $B_\bot$, $\delta g(B_\bot, 0)$, for various
temperatures with $B_\parallel=0$ for the
$8.0\,\mathrm{\mu m^2}$ dot (squares), along with RMT fits
      (solid curves).
      (b) Spin-orbit lengths $\lambda_{so}$ (circles) and phase
coherence times $\tau_{\varphi}$
      (triangles) as a function of temperature, from data in (a).
      \vspace{-0.2in}
      }}
\end{figure}

Besides $\epsilon_Z$ (which is calculated using $g = -0.44$ rather than fit), parallel field combined with SO coupling
introduces an additional new energy scale, $\epsilon^Z_\perp = {\kappa_z\epsilon_Z^2 A\over 2E_{T}}\sum_{i,j = 1,2}{l_i
\over \lambda_i}{l_j \over \lambda_j}$, where $\kappa_Z$ is a dot-dependent constant and $l_{1,2}$ are the components
of a unit vector along $B_\parallel$ \cite{RMT}. Because orbital effects of $B_\parallel$ on $\delta
g(B_\perp,B_\parallel)$ dominate at large $B_\parallel$, $\epsilon^Z_\perp$ must instead be estimated from RMT fits of
$\mathrm{var}(g)$ with already-broken time reversal symmetry, which is unaffected by orbital coupling
\cite{Zumbuhlvar}.

The RMT formulation \cite{RMT} is invariant under $\nu_{so}\rightarrow r/\nu_{so}$, where $r=L_1/L_2$ \cite{sym}, and
gives an extremal value of $\delta g(0,B_\parallel)$ at $\nu_{so}=\sqrt{r}$.  As a consequence, fits to $\delta
g(0,B_\parallel)$ cannot distinguish between $\nu_{so}$ and $r/\nu_{so}$. As shown in Fig.~2(b), data for the $8
\mathrm{\mu m^2}$ dot ($r\sim 2$) are consistent with $1 \leq \nu_{so} \leq 2$ and appear best fit to the extremal
value, $\nu_{so}\sim 1.4$. Values of $\nu_{so}$ that differ from one indicate that both Rashba and Dresselhaus terms
are significant, which is consistent with 2D data taken on the same material \cite{Miller}.

Using $\nu_{so}=1.4$ and values of $\lambda_{so}$, $\tau_\varphi$, and $\kappa_\bot$ from the $B_\parallel=0$ fit, RMT
predictions for $\delta g (B_\bot, B_\parallel)$ agree well with experiment up to about $B_\parallel\sim\,
0.2\,\mathrm{T}$ (Fig.\ 2(a)), showing a crossover from AL to WL. For higher parallel fields, however, experimental
$\delta g$'s are suppressed relative to RMT predictions. By $B_\parallel \sim 2\,\mathrm{T}$, WL has vanished in all
dots (Fig.\ 2(c)) while RMT predicts significant remaining WL at large $B_\parallel$. The full range of $\delta
g(0,B_\parallel)$ for the three dots is shown in Fig. 2(c). The center-gated ($5.6\,\mathrm{\mu m^2}$) dot and the
small ($1.2\,\mathrm{\mu m^2}$) dot show WL for all $B_\parallel$, and a similar suppression of WL above $B_\parallel
\sim 2\,\mathrm{T}$.

One would expect WL/AL to vanish once orbital effects of $B_\parallel$ break time reversal symmetry. Following
Ref.~\cite{Falko} (FJ), we account for this with a suppression factor
$f_{FJ}(B_\parallel)=(1+\tau_{B\parallel}^{-1}/\tau_{esc}^{-1})^{-1}$, where $\tau_{B\parallel}^{-1}\sim
aB_\parallel^2+bB_\parallel^6$, and assume that the combined effects of SO coupling and flux threading by $B_\parallel$
can be written as a product, $\delta g(0,B_\parallel)=\delta g_{RMT}(0,B_\parallel) \cdot f_{FJ}(B_\parallel)$. The
$B_\parallel^2$ term reflects surface roughness or dopant inhomogeneities; the $B_\parallel^6$ term reflects the
asymmetry of the quantum well. We consider fits taking $a$ as a fit parameter ($a_1$, Table I) with $b=1.4\, 10^8\,
\mathrm{s^{-1}T^{-6}}$ fixed, obtained from self-consistent simulations \cite{FJsim}, or allowing both $a$ and $b$ to
be fit parameters ($a_2$ and $b_2$, Table I). Figure 2(c) shows that allowing both to be free is only significant for
the (unusually shaped) center-gated dot; for the small and large dots, the single-parameter ($a$) fit gives good
quantitative agreement.

We next consider the effects of temperature and dephasing. We find that increased temperature reduces the overall
magnitude of $\delta g$ and also suppresses AL compared to WL, causing AL at $300\,\mathrm{mK}$ to become WL by
$1.5\,\mathrm{K}$ (maximum of $\delta g(B_\bot,0)$ at $B_\bot=0$ becomes minimum) in the $8\,\mathrm{\mu m^2}$ dot
(Fig.\ 3a). Fits of RMT to $\delta g(B_\perp,0)$ yield $\lambda_{so}$ values that are roughly independent of
temperature (Fig.\ 3b), consistent with 2D results \cite{Millo}, and $\tau_{\varphi}$ values that decrease with
increasing temperature. Dephasing is well described by the empirical form $(\tau_{\varphi}\mathrm{[ns]})^{-1}\sim
7.5\,\mathrm{T[K]} + 2.5\,\mathrm{(T[K])^2}$, consistent with previous measurements in low-SO dots
\cite{HuibersDephasing}. As temperature increases, long trajectories that allow large amounts of spin rotations are
being cut off by the decreasing $\tau_{\varphi}$ and the AL peak is diminished, as observed.

\begin{figure}[t]
             \label{figp4.epsf}
             \includegraphics{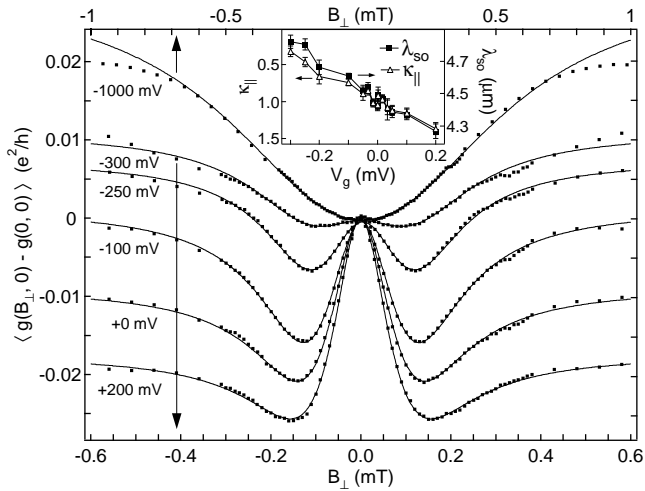}
      \vspace{-0.3in}
       \caption{\footnotesize{Difference of average conductance
$\langle g \rangle$ from its value at $B_\bot = 0$ as a function of $B_{\bot}$ for various center gate voltages $V_g$
in the center-gated dot (squares), along with fits to RMT \cite{RMT}. Good fits are obtained though the theory assumes
homogeneous SO coupling. Error bars are the size of the squares. Inset: $\lambda_{so}$ and $\kappa_\parallel$ as a
function of $V_g$ extracted from RMT fits, see text.
      \vspace{-0.2in}
       }}
\end{figure}

Finally, we demonstrate \emph{in situ} control of the SO coupling using a center-gated dot. Figure 4 shows the observed
crossover from AL to WL as the gate voltage $V_g$ is tuned from $+0.2\,\mathrm{V}$ to $-1\,\mathrm{V}$. At $V_g =
-1\,\mathrm{V}$, electrons beneath the center gate are fully depleted producing a dot of area $5.8\,\mathrm{\mu m^2}$
which shows WL. In the range of $V_g\geq-0.3$ V, the region under the gate is not fully depleted and the amount of AL
is controlled by modifying the density under the gate. Note that for $V_g>0\,\mathrm{V}$ the AL peak is larger than in
the ungated $8\,\mathrm{\mu m^2}$ dot. We interpret this enhancement not as a removal of the SO suppression due to an
inhomogeneous SO coupling \cite{BrouwerInho}, which would enhance AL in dots with $L/\lambda_{so} \ll 1$ (not the case
for the $8\,\mathrm{\mu m^2}$ dot), but rather as the result of increased SO coupling in the higher-density region
under the gate when $V_g>0\,\mathrm{V}$.

One may wish to use the evolution of WL/AL as a function of $V_g$ to extract SO parameters for the region under the
gate. To do so, the dependence may be ascribed to either a gate-dependent $\lambda_{so}$ or to a gate-dependence of a
new parameter $\kappa_\parallel=\epsilon_\parallel^{so}/(((L_1/\lambda_1)^2+(L_2/\lambda_2)^2)\epsilon_\bot^{so})$.
Both options give equally good agreement with the data (fits in Fig.~4 assume $\lambda_{so}(V_g)$), including the
parallel field dependence (not shown). Resulting values for $\lambda_{so}$ or $\kappa_\parallel$ (assuming the other
fixed) are shown in the inset in Fig.~4. We note that the 2D samples from the same wafer did not show gate-voltage
dependent SO parameters \cite{Miller}. However, in the 2D case a cubic Dresselhaus term that is not included in the RMT
of Ref.~\cite{RMT} was significant. For this reason, fits using \cite{RMT} might show $\lambda_{so}(V_g)$ though the 2D
case did not. Further investigation of the gate dependence of SO coupling in dots will be the subject of future work.

We thank I. Aleiner, B. Altshuler, P. Brouwer, J. Cremers, V. Falko, J. Folk, B. Halperin, T. Jungwirth and Y.
Lyanda-Geller. This work was supported in part by DARPA-QuIST, DARPA-SpinS, ARO-MURI and NSF-NSEC. Work at UCSB was
supported by QUEST, an NSF Science and Technology Center. JBM acknowledges partial support from NDSEG.


{

\begin{thebibliography}{10}
\bibitem{Subb}
\bibnamefont{ All measured densities are below the threshold for second subband occupation $n\sim 6.6\times 10^{15}\,
\mathrm{m^{-2}}$, which is known from Shubnikov-de Haas measurements and a decreasing mobility with increasing density
near the threshold. }
\bibitem{sym}
\bibnamefont{
The symmetry is precise if one takes $\epsilon_{\bot}^Z = \kappa_z \frac{\epsilon_Z^2}{2 E_T} {A \over
\lambda_{so}^2}$. See Ref.~\cite{RMT}. }
\end{thebibliography}
 }
\end{document}